\documentclass{emulateapj}
\usepackage{graphicx,amsmath,amsfonts,amssymb,subfigure,natbib,epsfig}

\def\spose#1{\hbox to 0pt{#1\hss}}
\def\simlt{\mathrel{\spose{\lower 3pt\hbox{$\mathchar"218$}}
     \raise 2.0pt\hbox{$\mathchar"13C$}}}
\def\simgt{\mathrel{\spose{\lower 3pt\hbox{$\mathchar"218$}}
     \raise 2.0pt\hbox{$\mathchar"13E$}}}

\shorttitle{Local Group dE Direct Ages}
\shortauthors{Geha~et~al.}

\begin{document}


\title{HST/ACS Direct Ages of the Dwarf Elliptical Galaxies NGC 147 and NGC 185}


\author{M.\ Geha\altaffilmark{1}}

\author{D.\ Weisz\altaffilmark{2,3}}

\author{A.\ Grocholski\altaffilmark{4}}

\author{A.\ Dolphin\altaffilmark{5}}

\author{R.\ P.\ van der Marel\altaffilmark{6}} 

\author{P.\ Guhathakurta\altaffilmark{7}}

\altaffiltext{1}{Astronomy
  Department, Yale University, New Haven, CT~06520.
  marla.geha@yale.edu}

\altaffiltext{2}{Astronomy Department, Box 351580, University of Washington, Seattle, WA}

\altaffiltext{3}{Hubble Fellow}

\altaffiltext{4}{Department of Physics \& Astronomy, Louisiana State University, Baton Rouge, LA 70803}

\altaffiltext{5}{Raytheon, 1151 E. Hermans Road, Tucson, AZ 85756, USA}

\altaffiltext{6}{Space Telescope
   Science Institute, 3700 San Martin Drive, Baltimore, MD~21218}

\altaffiltext{7}{UCO/Lick Observatory, University of California,
   Santa Cruz, 1156 High Street, Santa Cruz, CA~95064.}


\begin{abstract}
\renewcommand{\thefootnote}{\fnsymbol{footnote}}

We present the deepest optical photometry for any dwarf elliptical
(dE) galaxy based on {\it Hubble Space Telescope} ACS observations of the
Local Group dE galaxies NGC 147 and NGC 185. The resulting F606W and
F814W color-magnitude diagrams are the first to reach below the main
sequence turnoff in a dE galaxy, allowing us to determine full star
formation histories in these systems. The ACS fields are located
$\sim$1.5 effective radii from the galaxy center to avoid photometric
crowding. While our ACS pointings in both dEs show unambiguous
evidence for old and intermediate age stars, the mean age in NGC 147
is $\sim4$ Gyr younger as compared to NGC 185. In NGC 147, only 40\%
of stars were in place 12.5 Gyr ago (z$\sim$5), with the bulk of the
remaining stellar population forming between 5 to 7 Gyr.  In
contrast, 70\% of stars were formed in NGC 185 field more than 12.5
Gyr ago, with the majority of the remaining population forming between
8 to 10 Gyr. Star formation ceased in our ACS fields in both dEs
at least 3 Gyr ago. Previous observations in the central regions of
NGC 185 show evidence for star formation as recent as 100 Myr ago and
a strong metallicity gradient with radius. We suggest that the orbit
of NGC 185 has a larger pericenter as compared to NGC 147, allowing it
to preserve radial gradients and maintain a small central reservoir of
recycled gas. We interpret the inferred differences in star formation
histories to imply an earlier infall time into the M31 environment for
NGC 185 as compared to NGC 147.

\end{abstract}




\section{Introduction}\label{intro_sec}
\renewcommand{\thefootnote}{\fnsymbol{footnote}}

%%%%%%%%%%% FIGURE 1 %%%%%%%%%%%%%
\begin{figure*}[t!]
%\plotone{ACS_dss.eps}
\plotone{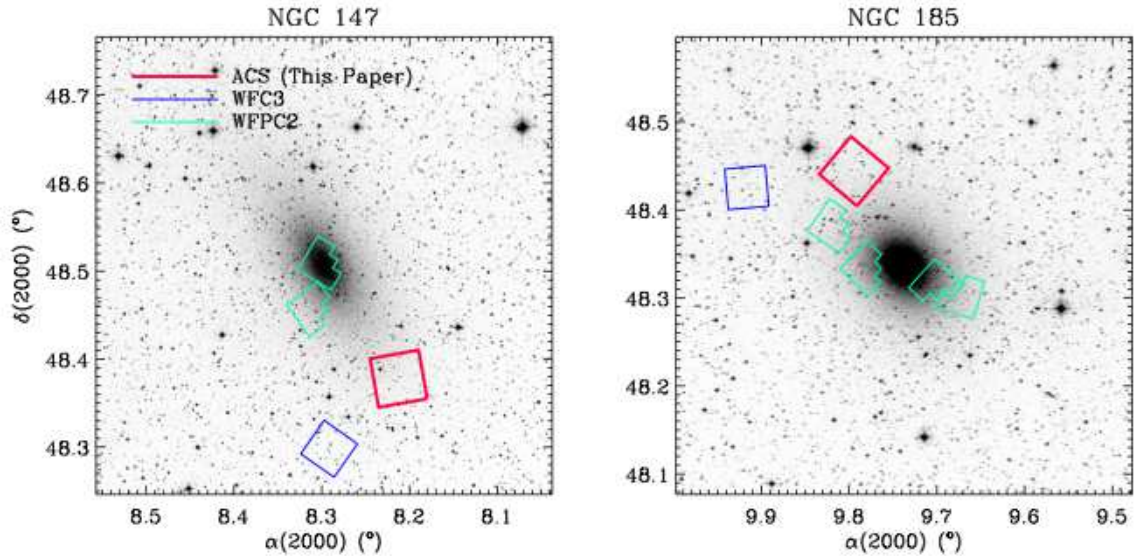}
\caption{Digital Sky Survey images of the Local Group dE galaxies,
  NGC~147 ({\it left}) and NGC~185 ({\it right}).  The ACS pointings
  discussed in this paper are shown in red.  WFC3 pointings taken in
  parallel with the ACS data are shown in blue and will be analyzed in
  a future paper.  Archival {\it HST}/WFPC2 pointings are shown in
  green. Images are $30'\times 30'$; North is up, East is to the
  left. \label{fig_dss}}
\end{figure*}

%  N147  RA = 8.300500   DEC = 48.508750
%  N185 =  RA 9.741542   DEC = 48.337389
%  Luis:   N147  -0.5 +/- 0.1   0.3 +/-0.1   basically flat fe/h
%              N185  -0.9 +/- 0.1   0.1 +/- 0.1  decreasing fe/h w/radius  

Dwarf elliptical (dE) galaxies are primarily, if not exclusively,
found in galaxy cluster and group environments \citep{binggeli88a,
  ferguson91a, geha12a}.  Environment-driven processes must play an
important role in dE formation and evolution \citep[e.g.,][]{moore98a,
  mastropietro05a, mayer10a,lisker13a}.  However, no single process or
formation mechanism can yet explain the observed diversity of dE
morphologies \citep[e.g.,][]{lisker09a, Janz12a} or kinematics
\citep[e.g.,][]{geha02a, vanzee04a, toloba14a}.  Star formation
histories are an important tool in understanding the dE galaxy class
and provide insight into the assembly of galaxies in dense environments.

Stellar population studies in the nearby Virgo and Fornax galaxy
clusters reveal a mixture of old to intermediate age stars in dE
galaxies, and intrinsic differences in the age and/or metallicity of
individual galaxies \citep{geha03a,michielsen08a, rys13a}.  These
studies are based on either integrated spectral line widths or narrow
band imaging and cannot therefore determine the exact ratio of old to
intermediate stars, nor reconstruct detailed star formation histories.
The Local Group dEs share the same broad properties as galaxy cluster
dEs \citep{bender91a}, yet are sufficiently nearby that individual
stars can be resolved below the main sequence turn-off, allowing for
more unambiguous stellar population estimates in these systems.

The Local Group contains three dE galaxies: NGC~205, NGC~147 and
NGC~185, all satellites of the large spiral galaxy M\,31.  A fourth
bright satellite, M\,32, is far more compact and likely has a
very different formation path \citep{monachesi12a}.  Of the three dEs, NGC~205 is close in
projection to the M\,31 disk, and contamination from M\,31 stars complicates
stellar population work in this object \citep{choi02a}.  The stellar
populations of NGC~147 and NGC~185 are more amenable to study.  RR
Lyrae stars are detected in both NGC~147 and NGC~185 indicative of
stars older than 10\,Gyr \citep{saha90a, yang10a}.  AGB carbon stars
are also observed in both dEs, typical of more intermediate age stars
\citep{battinelli04a,davidge05a,sohn06b}.   Each of these galaxies hosts its
own system of associated globular clusters \citep{Veljanoski13a}.

NGC~147 and NGC~185 share fundamental properties, such as absolute
luminosity \citep[$M_V\sim -16$;][]{pandas14a} and velocity dispersion
\citep[25 km/s;][]{geha10a}.  However, they differ in many aspects.
NGC~185 contains some gas, dust and evidence for recent star formation
confined to its center, while NGC~147 is devoid of gas or dust and
shows no sign of recent star formation activity \citep{young97a,
  marleau10a}.  {\it HST}/WFPC2 images of both galaxies also imply the
presence of intermediate-age stars, with NGC~147 having a more
significant contribution than NGC~185 based on stars brighter than the
main sequence \citep{butler05a, weisz14b}.  Spectroscopic
metallicities show an overall more metal-rich population in NGC 147,
[Fe/H] = $-0.5\pm0.1$ as compared to NGC 185 of [Fe/H] = $-0.9\pm0.1$
\citep{vargas14a}.  Only NGC 185 shows evidence for a metallicity
gradient, becoming more metal-poor with radius \citep{vargas14a,
  pandas14a}

NGC 147 and NGC 185 are very close in projection on the sky ($58'$),
leading some to argue that they may be a bound galaxy pair
\citep{vandenbergh98a}.  However, kinematic evidence suggests that
these may not be gravitationally bound \citep{geha10a, Watkins13a}.
Furthermore, deep photometry from the Pan-Andromeda Archaeological
Survey (PAndAs) has uncovered isophotal twisting at
large radius due to the emergence of symmetric tidal tails in NGC 147,
which is not seen to similar depths in NGC 185 \citep{pandas14a},
further suggesting these two objects have independent formation
paths.

In this paper, we present deep {\it Hubble Space Telescope (HST)}
Advanced Camera for Surveys (ACS) imaging for NGC~147 and NGC~185.
These are the only two dE galaxies for which {\it HST} can cleanly
resolve stars below the main sequence turnoff region and thus directly
measure their star formation histories.  The paper is organized as
follows: in \S\,\ref{sec_data} we discuss the {\it HST} observations
and data reduction.  We first compare the
resulting color-magnitude diagrams to  single
stellar population models (\S\,\ref{ssec_simple}), and then compute full star
formation histories (\S\,\ref{ssec_full}).  In \S\,\ref{sec_disc}, we discuss our
results in the context of dE galaxy formation.

%%%%%%%%%%%%%%%%%%%%%%%%%%%%%%%%%%
% Figure: CMD NGC 147
%%%%%%%%%%%%%%%%%%%%%%%%%%%%%%%%%%
\begin{figure*}[t]
\epsscale{1.1}
%\plottwo{cmd_scat1.eps}{cmd_scat2.eps}
\plotone{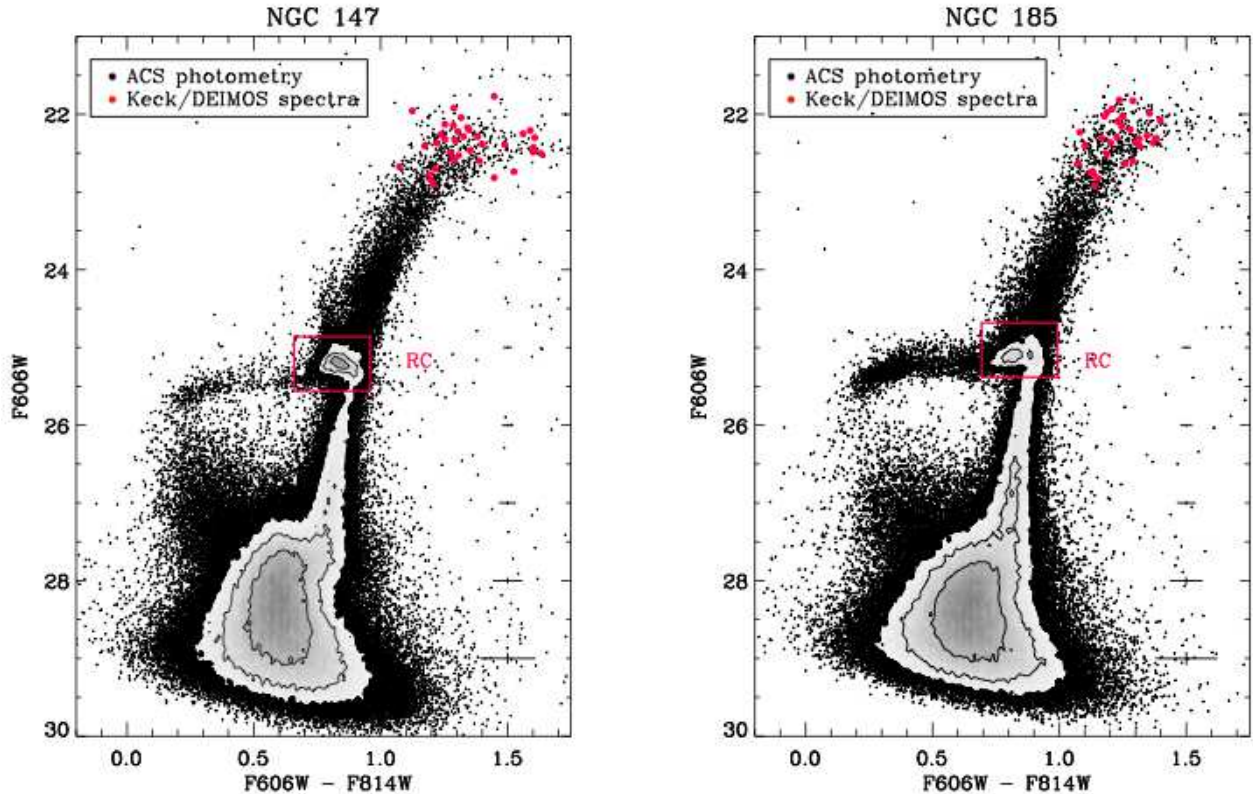}
\caption{Color-magnitude diagrams based on {\it HST}/ACS data for
  NGC~147 ({\it left}) and NGC~185 ({\it right}).  One sigma error
  bars are indicated to the right of each CMD.  Red symbols show the
  position of member stars in each ACS field with Keck/DEIMOS
  spectroscopic metallicity measurements.  While initially these two
  CMDs appear similar, closer inspection shows differences in the Red
  Clump (RC), horizontal branch and main sequence turnoff
  regions. \label{fig_cmd}}
\end{figure*}

\section{Data}\label{sec_data}

\subsection{Image Reduction and Photometry}

The {\it HST}/ACS observations were taken for NGC~147 between 2009
November-December, and for NGC~185 between 2010 January-February (HST
GO-11724; PI: Geha).  We note that this project was first approved as
HST-GO-10794 (PI: Geha), but was not completed under this program
number after the 2007 ACS failure.  Our post-servicing mission ACS
fields were chosen to avoid crowding while ensuring a well-populated
CMD in the F606W (broad $V$) and F814W ($I$) filters, corresponding to
a surface brightness of $\mu_V \sim 26$ mag\,arcsec$^{-2}$.  The exact
field locations were chosen to avoid bright foreground stars at this
radius.  Parallel WFC3 observations were taken as part of this
program.  A quick reduction of the WFC3 fields suggest a much sparser
stellar density as expected, but similar stellar populations as the
primary ACS fields; these data will be analyzed in a future
contribution.

The ACS fields are located roughly along the major axis of each galaxy
(Figure~\ref{fig_dss}), at $8.6'$ (1.8\,kpc) and $6.7'$ (1.2\,kpc)
from the galaxy center for NGC 147 and NGC 185, respectively.  Using
the effective radii determined by \citet{pandas14a} via deep CFHT/PAndAs
imaging, this corresponds to 1.3 and 2.3 effective radii from each
galaxy center.

The combined ACS exposure times were 82280\,s (23\,hours) for NGC~147
and 62598\,s (17\,hours) for NGC~185.  Individual exposures were well
dithered to fill in the gap between the two ACS chips and improve the
point-spread function (PSF).  Short 30-second exposures were taken as
part of the program, but were not included in our reduced image stacks
as bright stars are not the focus of this paper (individual long
exposures saturate brighter than $V\sim 20$).  The ACS field centers
and total exposure times are listed in Table~1.  The location of all
fields, including parallel WFC3 and archival WFPC2 pointings, are
shown in Figure~\ref{fig_dss}.

Our ACS images were processed through the {\it HST} archive on-the-fly
reprocessing system using the most up-to-date calibration frames.  The
images were corrected for CCD charge transfer efficiency
losses following the prescription in \citet{anderson10a}.  We then
co-added the images into a single image per field per filter using the
MULTIDRIZZLE software \citep{fruchter09a}, which also corrects for
geometric distortion and removes bad pixels and cosmic rays.  Due to
the multiply dithered images, the ACS images were resampled to a final
pixel size of $0.04''$ (0.8 times the native plate scale).  

PSF-fitting photometry was performed using the DAOPHOT/ALLSTAR II
package \citep{stetson94a}.  To create the PSF model we chose $> 400$
stars with good spatial coverage in each filter for each galaxy.
Nearby neighbors were subtracted from around the PSF stars and the
cleaned stars were then used to make the final PSFs.  To find faint
stars, we performed a single iteration of fitting the known stars,
subtracting them from the images and using this image to search for
faint companions.  We then re-performed the PSF fitting photometry on
the entire original image using the updated catalog.  Stars were
selected if they were found in both filters within a matching radius
of 0.5 pixels and passed the following criteria: $\sigma_{F606, F814}
< 0.15$\,mag, $\chi^2 < 3$ and abs(sharpness) $< 0.65$.  The number of
stars in each field is listed in Table~1.  Instrumental magnitudes
were converted into the ACS VEGAmag system using the procedure
outlined by \citet{sirianni05a}, but with updated ACS zeropoints
\citep{bohlin07a}.

Artificial star tests were performed by adding a total of a half
million artificial stars in each image stack.  Artificial stars were
spread evenly across the full frame and spaced such that they were
separated from each other by at least twice the PSF radius. PSF
fitting photometry was then performed on the entire frames, measuring
both the real and artificial stars, using the same steps as above.
Artificial stars were recovered if they passed the same photometric
criteria above.  The 50\% completeness levels are F606W = 29.05 and
28.95\,mag, and F814W = 28.35 and 28.25\,mag for NGC~147 and NGC~185,
respectively.

Contamination from either foreground Milky Way stars or stars
associated with the M\,31 halo is expected to be minimal in our ACS
fields.  Using the Besan\c con model of the Milky Way
\citep{robin03a}, we calculate there are $\sim40$ Milky Way stars
contained within each ACS field-of-view in the magnitude and color
range F814W$= 22 - 28$ and (F606W-F814W)$< 1.5$.  This is negligible
compared to the hundreds of thousands of stars detected in each dE
galaxy (see Table~1).  For the model fitting in \S\,\ref{ssec_full},
we do account for Milky Way foreground contamination using the
empirical models of \citet{dejong10a}.  The
contamination from M\,31 stars at a projected radius of 7$^{\circ}$ or
100\,kpc is less than that from the Milky Way foregrounds and
we do not account for this in the models.

The resulting {\it HST} ACS CMDs are shown for NGC~147 and NGC~185 in
Figure~\ref{fig_cmd}.  We plot one sigma errors as a function of
magnitude.  These represent the deepest CMDs for any dE galaxy and are
the first observations of dE galaxy stars below the main sequence
turnoff.

%%%%%%%%%%%%%%%%%%%%%%%%%%%%%%%%%%
% Figure: CMD NGC 185
%%%%%%%%%%%%%%%%%%%%%%%%%%%%%%%%%%
\begin{figure*}[t]
\plottwo{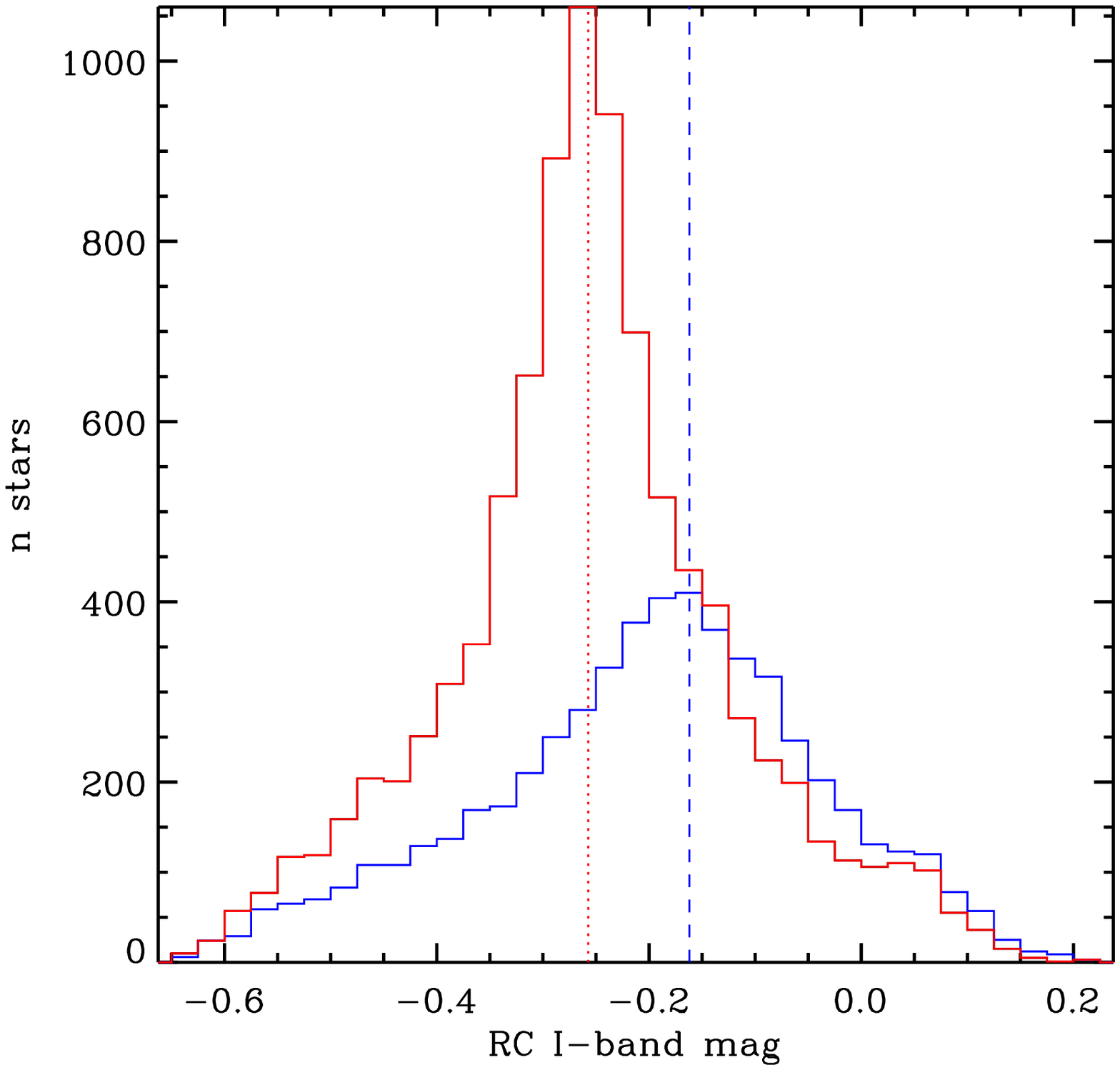}{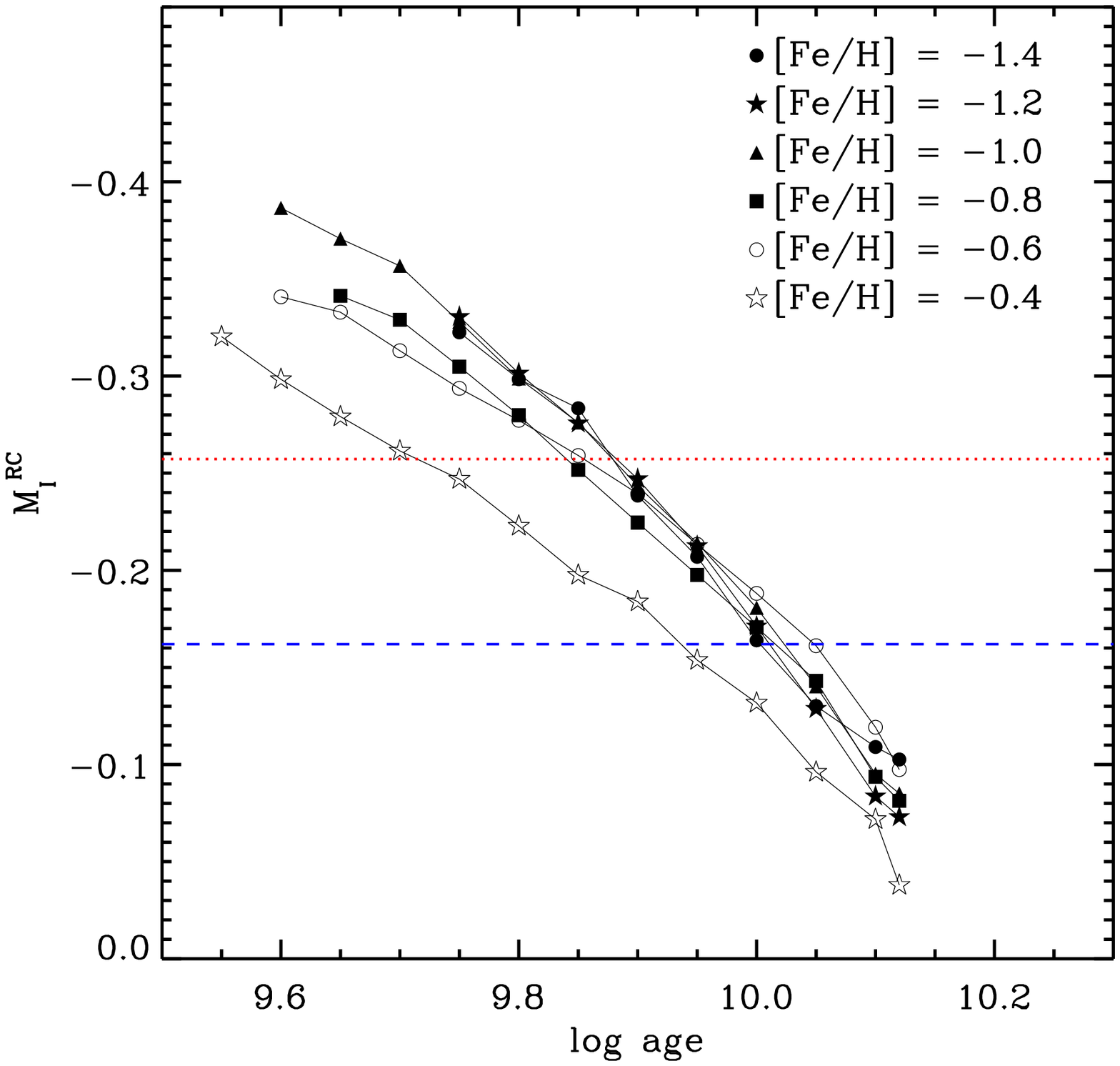}
\caption{As a first comparison of the SFHs, we compare the observed
  distribution of stars near the Red Clump with single isochrone
  models.  The red clump region of the CMD is shown inside the red box of
  Figure~\ref{fig_cmd}.  The peak of this distribution in absolute magnitude
  ({\it left}) occurs 0.1 magnitudes brighter for NGC~147 (red) as compared
  to NGC~185 (blue).  Comparing to isochrone predictions ({\it right}) for a
  range of metallities ([Fe/H] between $-0.4$ to $-1.4$), this implies
  a younger mean age for NGC~147. \label{fig_rc_age}}
\end{figure*}

\subsection{Distances and Spectroscopic Metallicities}
\label{ssec_dist_z}

%                        M31                    N147                  N185
% McConn       785+25-25           675+-27            616+-26  
%  Conn         779 +19-18        712+21/019       620+19-18 
% AG                                           724+-27            636 +-26

Given the extremely small photometric errors in our data, we chose to
re-determine the distance of NGC 147 and NGC 185 using the {\it
  HST}/ACS data itself via the Tip of the Red Giant Branch (TRGB)
method.  The TRGB acts as a discontinuity in the stellar luminosity
function, the position of which is easily measured.  We measure the
TRGB using the software developed by one of us (R.P.v.d.M.) and
detailed in \citet{cioni00a}.  The distance modulus for NGC~147 is
$(m -M)_0= 24.30 \pm 0.05$ ($724\pm 27$\,kpc) and NGC~185 is $(m -M)_0
= 24.02 \pm 0.08$ ($636\pm 26$\,kpc).  These are slightly further, but
well within the 1-$\sigma$ errors, of those determined by
\citet{Conn12a}.  We assume a distance to M\,31 of 779\,kpc from
\citet{Conn12a}.  This places NGC~147 and NGC~185 at distances 165 and
210\,kpc away from their parent galaxy M\,31, respectively.  We adopt
extinction values for NGC~147 and NGC~185 from \citet{schlegel98a} of
E(B-V) = 0.161 and 0.195.

Spectroscopic metallicities for stars in the ACS fields were obtained
as part of a larger Keck/DEIMOS study of Local Group dEs
\citep{geha10a, vargas14a,Ho15a}.  We use these spectroscopic
metallicities to guide our single stellar population analysis in
\S\,\ref{ssec_simple}.  The Keck sample includes 37 member stars
within our ACS field-of-view in NGC~147, and 30 member stars in the
NGC~185 ACS field.  Members are identified as having colors,
magnitudes and radial velocities consistent with each galaxy.
Spectroscopic metallicities ([Fe/H]) of individual RGB stars were
determined using the spectral synthesis method described in \citet{vargas14a}.
The mean metallicity within each ACS field is $<[{\rm Fe/H}]>=
-0.5\pm0.1$ for NGC~147 and $<[{\rm Fe/H}]>= -1.0\pm0.1$ for NGC~185,
comparable to the means of each full sample.  While the error on the
mean is small, there is significant internal scatter of 0.5\,dex in
each field.  While the analysis below assumes solar alpha-abundance
ratios, we note that the measured alpha-abundances are slightly
enhanced at $[\alpha/{\rm Fe}] = 0.3 \pm 0.1$ in NGC 147 and
$[\alpha/{\rm Fe}] =
0.1 \pm 0.1$ in NGC185 \citep{vargas14a}.

\section{Results}\label{sec_results}

In Figure~\ref{fig_cmd}, the {\it HST}/ACS CMDs for NGC 147 and NGC
185 appear similar at first glance, yet clear differences emerge on
closer inspection.  Focusing first on the position of stars which have
evolved off the main sequence, NGC~147 has a strong Red Clump (RC),
while stars in the same evolutionary state in NGC~185 are instead
distributed equally between the RC and a bluer horizontal branch,
consistent with a lower average metallicity in NGC~185.  Also seen in
the CMDs of Figure~\ref{fig_cmd} is the larger number of stars bluer
and brighter than the main sequence turnoff in NGC~147 as compared to
NGC~185.  While some fraction of these stars are most likely blue
straggler stars \citep{santana13a}, this does suggest the presence of
a younger population in NGC 147 as compared to NGC 185.  Based on this
visual inspection alone, we conclude that the star formation histories
of these two dEs are different.  Below, we quantify this statement via
both simple comparison to single population isochrones
(\S\,\ref{ssec_simple}) and CMD fitting to determine the full star
formation histories (\S\,\ref{ssec_full}).

%%%%%%%%%%%%%%%%%%%%%%%%%%%%%%%%%%
% Figure: FULL SFH
%%%%%%%%%%%%%%%%%%%%%%%%%%%%%%%%%%
\begin{figure*}[t]
\plotone{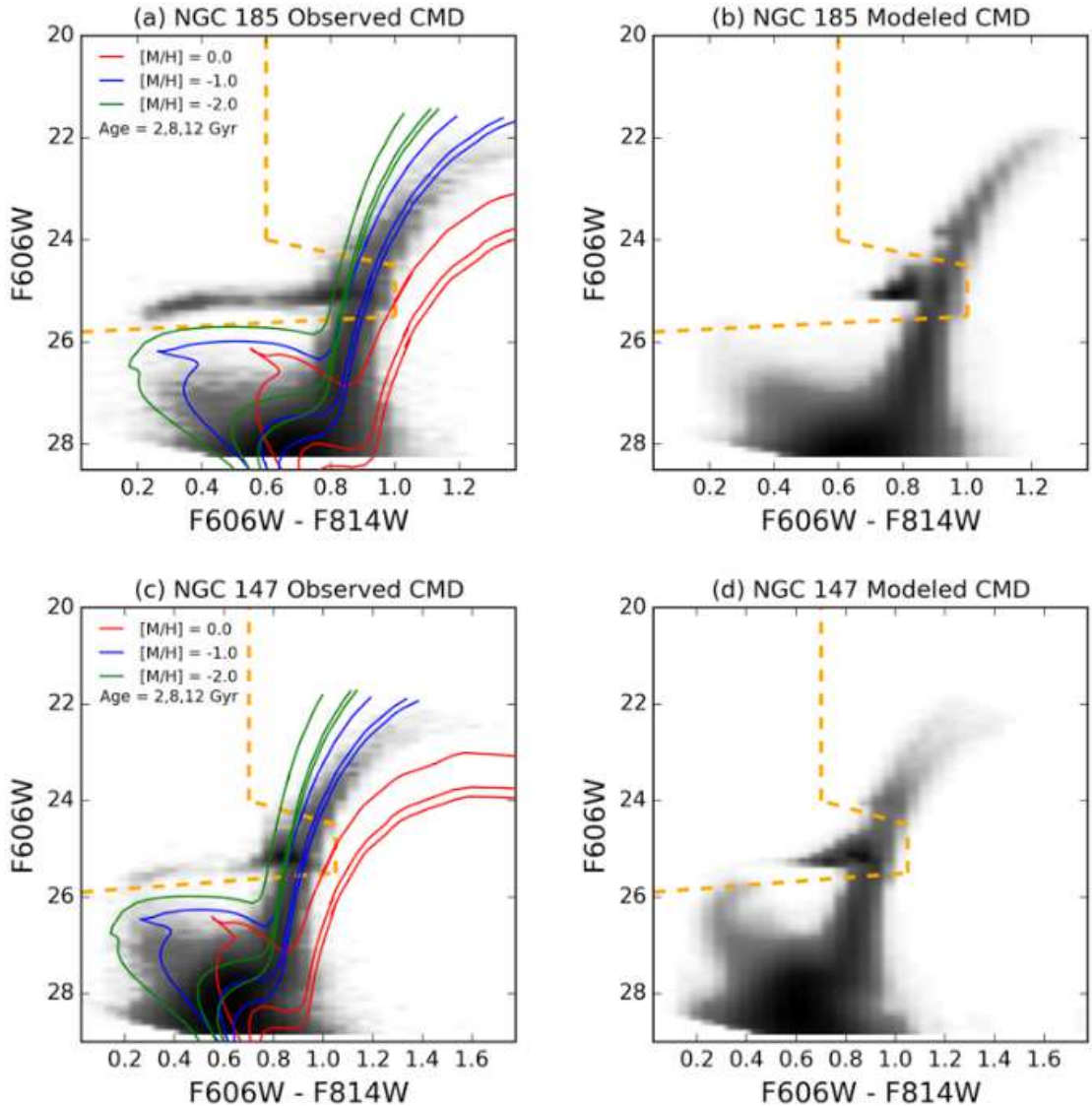}
\caption{{\it Left panels:\/} The observed Hess diagrams for NGC 185
  ({\it top}) and NGC 147 ({\it bottom}).  Overplotted are Padova
  isochrones with [Fe/H] = $-2$ (green), $-1$ (blue) and 0.0\,dex
  (red) for ages of 2, 8 and 12\,Gyr.  ({\it Right panels:\/}) The best fitting model
  star formation histories as determined using MATCH.  The region 
  above and to the left of the orange boundaries are not used in the
  CMD fitting.\label{fig_fullsfh}}
\end{figure*}

\subsection{Simple Constraints on the Distribution of Stellar Ages}
\label{ssec_simple}

Before modeling the inferred star formation histories from the full
CMDs, we first compare the data to single stellar population
isochrones to build a basic understanding of these data.  We use the Padova
isochrones \citep{girardi02a, girardi10a} generated in the {\it HST}
ACS filters with solar abundance ratios.

We focus on the relative ages of NGC~147 and NGC~185 by comparing
stars in the Red Clump (RC).  The location of the RC is shown in
Figure~\ref{fig_cmd}.  For metallicities more metal-poor than
$\sim\,-0.5$\,dex, the F814W (roughly $I$-band) luminosity of the RC
is largely dominated by age effects.  In the left panel of
Figure~\ref{fig_rc_age}, we plot the F814W luminosity functions of RC
stars in NGC 147 (red) and NGC 185 (blue).  NGC 147's median RC
magnitude is marked by the red dotted vertical line, while the blue dashed line
marks NGC 185's RC magnitude.  In the right panel of
Figure~\ref{fig_rc_age} we compare the median RC luminosity to the
values predicted for single age/metallicity stellar populations.  These values
were determined using the same RC boxes as the observations.

Spectroscopic abundances for NGC 147 and NGC 185 are [Fe/H] = $-0.5$
and $-1.0$, respectively (\S\,\ref{ssec_dist_z}), thus we can use the
RC to determine the relative ages of the two galaxies.  The dominant
RC population in NGC 147 has an age of $\sim$ 7\,Gyr, while NGC 185's
is $\sim$11\,Gyr old.  This makes NGC 185 about 4\,Gyr older. 
This is in agreement with the full SFHs discussed in
\S\,\ref{ssec_full} and serves an as independent check of those
result, as the full SFHs are determined without including this region of the CMD.

In the left panels of Figure~\ref{fig_fullsfh}, we overplot onto the
CMD Hess diagrams representative isochrones spanning a wide range of
metallicities (${\rm [Fe/H]} =0,-1~{\rm and}\, -2$).  For the best
fitting set of isochrones which fits the red giant branch (blue lines
for both dEs), the bulk of main sequence turn-off stars lie between
the 2 and 8\,Gyr tracks for NGC 147, while these stars primarily lie
between the 8 to 12\,Gyr tracks for NGC 185.  Finally, if all of the
stars bluer and brighter than the main sequence turnoff are
interpreted as young stars (some fraction must be older blue straggler
stars), than this population is likely not younger then $\sim2$\,Gyrs.
We explore this in more detail below.

%%%%%%%%%%%%%%%%%%%%%%%%%%%%%%%%%%
% Figure: Culmulatiev SFH
%%%%%%%%%%%%%%%%%%%%%%%%%%%%%%%%%%
\begin{figure*}[t]
\plotone{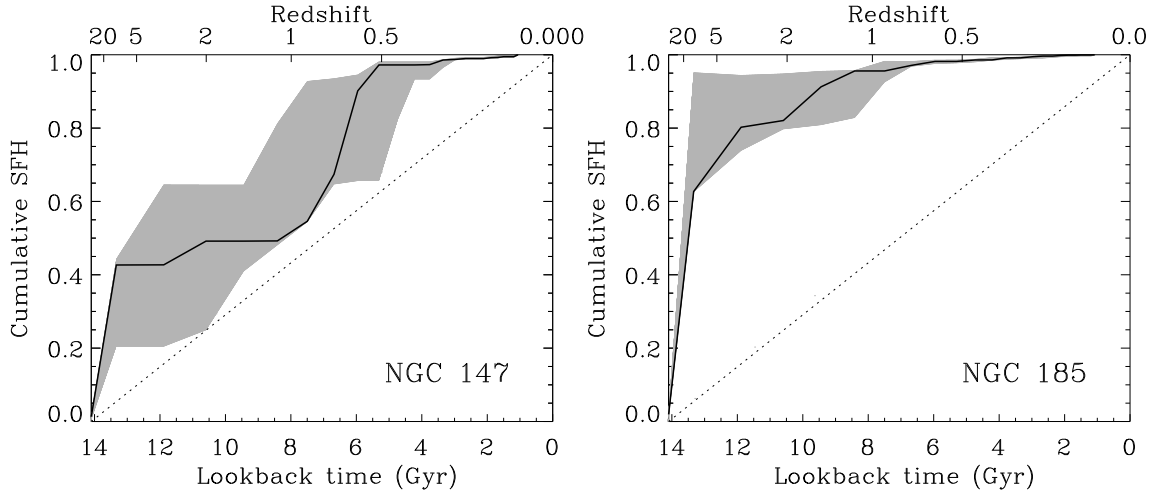}
\caption{Cumulative SFHs for NGC 147 ({\it left}) and NGC 185 ({\it
    right}) from the best-fitting SFHs models shown in
  Figure~\ref{fig_fullsfh}.  The shaded envelope reflects the 68\%
  confidence interval around the best fit SFH.  The dotted line
  indicates a constant star formation rate over the lifetime of the
  galaxy.  While the NGC 185 field is primarily old, the bulk of stars
  in NGC 147 formed between 5-7 Gyr ago. \label{fig_csfh}}
\end{figure*}

\subsection{Measuring Full Star Formation Histories}

We measure the full SFHs of both galaxies using the CMD fitting
package MATCH \citep{dolphin02a}.  MATCH constructs a set of synthetic
simple stellar populations which are linearly combined and added to a
model foreground population from the empirical model in
\citet{dejong10a} to form a composite synthetic CMD.  This CMD is
then convolved with results from artificial star tests.  The models
and observed CMDs are compared using a Poisson likelihood statistic.
The SFH that corresponds to the best matched synthetic CMD is the most
likely SFH of the observed population. The fitting of these CMDs uses the
Padova stellar models \citep{girardi02a, girardi10a} and follows the
fitting methodology from \citet{weisz14b}.  We assume a Kroupa IMF,
solar-scaled isochrones and the distances/reddening stated in
\S\,\ref{ssec_dist_z}.  We exclude the red clump and horizontal branch
regions from the fit in order to mitigate the contribution of these
relatively less certain phases of stellar evolution to the SFH
\citep{aparicio09a}.  The excluded region is shown above and to the
left of the orange dashed lines in Figure~\ref{fig_fullsfh}

Uncertainties in the SFHs reflect the 68\% confidence interval around
the best fit SFH due to both random uncertainties (from a finite
number of stars on the CMD) and systematic uncertainties (due to
uncertain physics in the underlying stellar models).   In both our ACS
datasets, systematic errors dominate over random.    We refer the
reader to \citet{dolphin12a} for a full discussion of systematic
uncertainties and \citet{dolphin13a} for a detailed description of random
uncertainties in SFH measurements.

\subsection{Full SFH Results}
\label{ssec_full}

The best-fitting MATCH CMD models are compared to the observed Hess
diagrams in Figure~\ref{fig_fullsfh}.  The cumulative star formation
histories from these models and their uncertainties are shown in
Figure~\ref{fig_csfh}.  The main features of the main sequence, main
sequence turnoff and upper red giant branches are generally well fit.  We note
minor discrepancies  between the model and data, particularly near the
bright end of the youngest main sequence turnoff in NGC 147.  This
population represents $\sim1$\% of the total stellar mass and may
be due to our assumption of solar alpha-abundance ratios or some other
unaccounted for systematic.  The cumulative SFHs of NGC~147 and
NGC~185 required to achieve the overall agreement are significantly
different from each other and are consistent with the general
conclusion from \S\,\ref{ssec_simple} that NGC 185 hosts a much older
stellar population as compared to NGC 147.

In NGC~147, less than half of stars (40\%) were in place 
12.5\,Gyrs ago ($z\sim5$), with the bulk of the remaining population
forming between 5 to 7\,Gyrs.  Star formation in this ACS field
appears to have been quenched roughly 5\,Gyrs ago.  This agrees
with SFHs determined from shallower {\it HST}/WFPC2 data from the more
central regions of NGC\,147 \citep{weisz14b}.  There are no
metallicity gradients observed in this galaxy
\citep{vargas14a,pandas14a}, suggesting that the inferred SFH in this
ACS field may be representative of the galaxy as a whole.
\citet{pandas14a} has recently uncovered isophotal twisting at large
radius due to the emergence of symmetric tidal tails in NGC 147.  Our
ACS field is well inside the radius at which these tidal effects are
visibly observed.

The NGC 185 field, in contrast, formed 70\% of stars prior to
12.5\,Gyrs ago, with the majority of the remaining population forming
between 8 to 10\,Gyrs ago.  Given NGC 185's lower measured
spectroscopic metallicity, an older age for NGC 185 relative to NGC
147 is in agreement with expectations from simple models of chemical
evolution \citep[e.g.,][]{tinsley79}.  We note that the spectroscopic
metallicities are not used as part of the fitting method.  The
presence of the smaller intermediate age (8-10\,Gyr) population is
independently confirmed by AGB carbon stars present in the ACS field
\citep{battinelli04a}.  A radial gradient in both carbon stars and
overall stellar metallicity is observed in NGC 185
\citep{vargas14a,pandas14a}, which suggests that NGC 185 has
experienced little to no radial mixing.  The star formation in the ACS
field, which lies at 2.3 effective radius from the galaxy center
($6.7'$ or 1.2\,kpc), is among the oldest of any Local Group dwarf
galaxy \citep{weisz14b}.  This is in contrast to the inner 200\,pc of
NGC 185 which shows evidence for very recent star formation as young
as 100\,Myrs ago.

\vskip 0.5cm

\section{Formation History of NGC147 versus NGC 185}\label{sec_disc}

Based on the star formation histories above, NGC~147 and NGC~185 built
up their present-day stellar mass over very different timescales.  A
key question is whether these galaxies had different formation
histories prior to falling into the M31 environment, or if their early
histories were similar, but each fell into the M31 potential at very
different times.  Below we argue that a different infall time into M31 is the
most plausible explanation for these different SFHs.

NGC147 and NGC 185 lie very close on the sky ($58'$) and have been
interpreted either as a bound galaxy pair or a single group that
fell together into M31 \citep{vandenbergh98a}.  Kinematic evidence marginally
suggests that these are not gravitationally bound \citep{geha10a,
  Watkins13a}.  Our star formation histories, combined with the
presence of a tidal tail in NGC 147 (and the lack thereof in NGC 185)
suggest that these two dE galaxies are independent of each other and that their
current proximity is likely a chance projection.  While current data
only support this conclusion indirectly, future proper motion
measurements can definitively determine whether or not these two
systems have different orbital histories.

An outstanding issue in comparing NGC 147 to NGC 185 is the difference
in gas and dust content.  NGC 185 contains some gas and dust ($M_{\rm
  gas} = 3\times 10^5 M_{\rm sun}$, $M_{\rm dust} = 5\times 10^3
M_{\rm sun}$, De Looze et al.~in prep., \citet{marleau10a}),
consistent with recycling from older stars.  NGC 147 contains no
detected gas nor dust ($M_{\rm gas} < 3\times 10^3 M_{\rm sun}$,
\citet{young97a}; $M_{\rm dust} < 128 M_{\rm sun}$, De Looze et al.~in
prep.).  The lack of gas/dust in NGC 147 has been referred to as the
'missing ISM problem' \citep{sage98a}.  We hypothesize that NGC 147
has had stronger interactions with M\,31 as compared to NGC 185 due to
an orbit which bring it deeper into the M31 potential, stripping
gas/dust and creating its observed tidal tails.  On the other hand, we
suggest that the orbit of NGC 185 has a larger comparable orbital
pericenter, allowing it to maintain a small gas reservoir at its
center.  We interpret the very recent star formation confined to the
central 200\,pc of NGC 185 as having formed from this recycled
material \citep{gonclaves12a}, which was perhaps triggered during a
recent passage through pericenter.

From Figure~\ref{fig_csfh}, we concluded that neither the NGC 147 nor
NGC 185 field have had a significant star formation event in several
gigayears (we do not consider the above 100\,Myr central burst in NGC
185 significant).  If we assume that star formation is
instantaneously quenched once a satellite passes the M31 virial radius, we
conclude infall times of 5 and 8\,Gyrs for NGC 147 and NGC 185
respectively.  We define 'quenched' as having 90\% of stars in place,
choosing this value (rather than 100\%) due to ambiguity with blue
straggler stars which are not accounted for in the models.  If there
is a delay time between infall and star formation quenching
\citep[e.g.,][]{wetzel13a}, then these times could be much earlier.
Because the majority of stellar mass in NGC 185 was in place earlier
than NGC 147, we conclude an earlier infall epoch for this galaxy
regardless of the quenching timescale.

Finally, we compare the cumulative SFHs of NGC 147 and NGC 185 to
other dwarf galaxy satellites around the Milky Way and M31 from
\citet{weisz14b}.  NGC 147 is typical of other luminous Local Group
satellites, with evidence for a mixture of both old (12\,Gyr) and more
intermediate age stars.  On the other hand, our NGC 185 field is more
typical of the predominately old populations seen in much lower
stellar mass systems such as Sextans and Draco.  In these lower mass
systems, star formation is assumed to have been suppressed due to
reionization.  However, NGC 185 has a factor of 100 more stellar mass
than these systems and unlikely to be affected to this extent by
reionization.  Instead this may be further evidence for NGC 185's
early infall time into the M31 environment.

\section{Conclusions} 

We present deep {\it HST} ACS photometry of the M31 dE satellite
galaxies NGC 147 and NGC 185.  These data are the first to reach below
the main sequence turnoff in a dE galaxy and allow us to unambiguously
determine their star formation histories.  We find a much older
population in the dE galaxy NGC 185 as compared to NGC 147: while 70\%
of stars were already formed in NGC 185 before 12.5\,Gyrs, only 40\%
of the present day stellar population had formed in NGC 147.  We
broadly interpret these results to imply an earlier infall time
into the M31 environment for NGC 185 versus NGC 147.   We further
suggest that the orbit of NGC 185 has a larger pericenter as
compared to NGC 147, allowing it to preserve radial age/metallicity
gradients and maintain a small central reservoir of recycled gas.   These conclusions
could be confirmed via future proper motion measurements of
these two systems to determine their 3D orbits around M31.

\vskip 0.2cm

\acknowledgments Support for this work was provided by NASA through
grant number HST GO-11724 and HST-GO-10794 from the Space Telescope
Science Institute, which is operated by AURA, Inc., under NASA
contract NAS5-26555.  We thank Ana Bonaca, Jeremy Bradford and Ilse de
Looze for comments on the draft.  MG acknowledges support from NSF
grant AST-0908752.  Support for DRW is provided by NASA through Hubble
Fellowship grant HST-HF-51331.01 awarded by the Space Telescope
Science Institute, which is operated by the Association of
Universities for Research in Astronomy, Inc., under NASA contract NAS
5-26555.  PG acknowledges support from NSF grant AST-1412648.

%% Make sure that you provide a unique KEY for every \bibitem or else they



\begin{deluxetable}{lcccc}
\tabletypesize{\scriptsize}
\tablecaption{HST ACS Observations}
\tablewidth{0pt}
\tablehead{
\colhead{Row} & \colhead{Quantity} & \colhead{Units} & \colhead{NGC
  147} & \colhead{NGC 185}
}
\startdata
(1) & RA            & h:m:s                      &  00:33:12.1    & 00:38:58.0 \\
(2) & DEC          & $^{\circ}: \> ': \> ''$ &  +48:30:31 & +48:20:15\\ 
(3) & E(B-V)       & mag                        & 0.161             &  0.195  \\
(4) & (M-m)$_0$ &                              &   $24.30\pm 0.05$     &   $24.02\pm0.08$   \\
(5) & Dist           & kpc                       &  $724\pm27$  & $636\pm 26$  \\
(6) & $M_{V,0}$   & mag                      & $-16.5\pm0.04$    & $-15.5\pm0.04$ \\
(7)  & $r_{\rm eff}$ & $'$                     & $6.7\pm0.09$     &   $2.94\pm0.04$\\
(8)  & $r_{\rm eff}$ & kpc                  & $1.41\pm0.02$ &   $0.53\pm0.01$\\ 
(9)  & [Fe/H]   & dex                       & $-0.5\pm0.1$     & $-0.9\pm0.1$   \\ \hline
(10) & RA$_{\rm HST/ACS}$                & h:m:s                      &  00:32:51.0    & 00:39:10.5 \\
(11) & DEC$_{\rm HST/ACS}$          & $^{\circ}: \> ': \> ''$ &  +48:22:39.8 & +48:26:35.9\\ 
(12)  &  t$_{\rm exp, F606W}$ & hrs                      & 10.65    & 12.2\\
(13) & t$_{\rm exp, F814W}$ & hrs                       & 7.54     & 9.84 \\
(14) & $N_{\rm stars}$ &                    & 277,949    & 197,634
\enddata
\tablecomments{Columns (1)-(3) are taken from NED for the galaxy
  center.   Columns (3) from \citet{schlegel98a}; column (4)-(5) are determined from our photometry as
  discussed in \S\,\ref{ssec_dist_z}.   Columns (6)-(8) are from
  \citet{pandas14a}; column (9) from \citet{vargas14a}.   The last five
  columns are specific to the {\it HST}/ACS data presented here.}
\end{deluxetable}


%% If you are not including electonic art with your submission, you may
%% mark up your captions using the \figcaption command. See the 
%% No more than seven \figcaption commands are allowed per page, 
%% so if you have more than seven captions, insert a \clearpage 

\end{document}